\documentclass{easychair}
\usepackage{makeidx}
\makeindex

\newcommand{\xf}[1]{Figure~\ref{#1}}

\newcommand{\xs}[1]{Section~\ref{#1}}

\newcommand{\xl}[1]{Listing~\ref{#1}}

\newcommand{\gee}{{GEE\index{GEE}\index{Frameworks!GEE}}}

\newcommand{\gipsy}{{GIPSY\index{GIPSY}}}

\newcommand{\gipl}{{GIPL\index{GIPL}}}

\newcommand{\lucid}{{Lucid\index{Lucid}}}

\newcommand{\jlucid}{{JLucid\index{JLucid}}}
\newcommand{\olucid}{{Objective Lucid\index{Tensor Lucid}}}

\newcommand{\flucid}{{Forensic Lucid\index{Forensic Lucid}}}

\newcommand{\lucx}{{Lucx\index{Lucx}}}

\newcommand{\jooip}{{JOOIP\index{JOOIP}}}

\newcommand{\java}{{Java\index{Java}}}

\newcommand{\lisp}{{LISP\index{LISP}}}

\newcommand{\lucidop}[1]{{\bf \texttt{#1}}}

\newcommand{\trans}{$\psi$}

\newcommand{\invtrans}{$\Psi^{-1}$}

\newcommand{\api}[1]{\texttt{#1}\index{API!#1}}

\newcommand{\lucidL}[1]{{$\mathit{Lucid}$}($L$) }

		{}

\def\myvert{\raise 2.27pt \hbox{\vrule depth 0pt height 8pt width 0.2mm}}
\def\myarrow{\hspace*{0.43mm}%
             \raise 2.29pt\hbox{\vrule depth 0pt height 8pt width 0.16mm}%
             \hspace*{-0.32mm}%
             $\longrightarrow$
             \ %
             }

\lstdefinestyle{codeStyle}
{
        language=Java,
        frame=single,  %use this to have a box around the code
        basicstyle=\scriptsize,
        captionpos=b,
        showstringspaces=false,
        showspaces=false,
        extendedchars=true,
        linewidth=1\linewidth,
        breaklines=true,
        float=htpb  %change this to change where the floats end
}

\begin{document}

\title{Reasoning About a Simulated Printer Case Investigation with Forensic Lucid}
\titlerunning{Printer Case Investigation with Forensic Lucid}

\author{Serguei A. Mokhov \hspace{1cm} Joey Paquet \hspace{1cm} Mourad Debbabi\\
Faculty of Engineering and Computer Science\\
Concordia University, Montr\'eal, Qu\'ebec, Canada,\\
\url{{mokhov,paquet,debbabi}@encs.concordia.ca}%
}

\authorrunning{Serguei A. Mokhov et al.}

\maketitle

\begin{abstract}
In this work we model the ACME (a fictitious company name) ``printer case incident'' and make its specification
in {\flucid}, a {\lucid}- and intensional-logic-based programming language for
cyberforensic analysis and event reconstruction specification. The printer
case involves a dispute between two parties that was previously solved using
the finite-state automata (FSA) approach, and is now re-done in a more usable
way in {\flucid}.
Our simulation is based on the said case modeling by encoding concepts like
evidence and the related witness accounts as an evidential statement context
in a {\flucid} program,
which is an input to the transition function that models the possible
deductions in the case. We then invoke the transition function (actually its
reverse) with the evidential statement context to see if the evidence we
encoded agrees with one's claims and then attempt to reconstruct the sequence
of events that may explain the claim or disprove it.
\end{abstract}

\tableofcontents
\listoffigures
\lstlistoflistings

\section{Introduction}

\subsection{Problem Statement}

The very first formal approach to 
cyberforensic analysis and event reconstruction
appeared in two papers~\cite{printer-case,blackmail-case}
by Gladyshev et al. that relies on the finite-state automata (FSA) and their transformation
and operation to model evidence, witnesses,
stories told by witnesses, and their possible evaluation for the purposes of
claim validation and event reconstruction.
One of the examples the papers present
is the use-case for the proposed technique -- the ``ACME Printer Case Investigation''.
See~\cite{printer-case}
for the corresponding formalization using the FSA by Gladyshev and the 
proof-of-concept {\lisp} implementation.
We aim at the same case to model and implement it
using {\flucid}, which paves a way to be more friendly and usable in the
actual investigator's work and serve as a basis to further development in the area.

\subsection{Proposed Solution}

We show the intensional approach to the problem is an asset in the field of cyberforensics as it
is promising to be more practical and usable than the plain FSA and {\lisp}.
Since {\lucid} was originally designed and used to prove correctness of programming
languages~\cite{lucid76,lucid77,lucid85,lucid95}, and is based on the temporal logic, functional and data-flow languages
its implementation to backtracking in proving or disproving the evidential statements and claims in the
investigation process as a evaluation of an expression that either evaluates to {\em true} or {\em false} given all
the facts in the formally specified context.
We will also try to retain the generality of the approach vs. building
a problem-specific FSA in the FSA approach that can suffer a state explosion problem.

\subsubsection{Intensional Logic}

From the logic perspective, it was shown one can model computations (the basic unit in the finite state machines
in~\cite{printer-case,blackmail-case}) as logic~\cite{lalement-plaice93}.
When armed with contexts as first-class
values and a demand-driven model
adopted in the implementation of the Lucid-family of
languages~\cite{gipsy2005,gipsy-simple-context-calculus-08,gipsy-multi-tier-secasa09,kaiyulucx,eager-translucid-secasa08,multithreaded-translucid-secasa08} that limits the scope of evaluation in a given set of
that constrains the scope of evaluation in a given set of
dimensions, we come to the intensional logic and the corresponding programming artifact.
In the essence, we model our forensic
computation unit in the intensional logic and
implement it in practice within an intensional programming
platform~\cite{gipsy-all-named,gipsy2005,mokhovmcthesis05}.
We project a lot of potential for this work to be successful, beneficial, and usable for
cyberforensics investigation
as well as simulation
and intensional programming communities.

\subsubsection{Approach Overview}

Based on the parameters and terms defined in the works of Gladyshev~\cite{printer-case,blackmail-case},
we have various pieces of evidence and witnesses telling their own ``stories'' of an incident.
The goal is to put them together to make the description of the incident as precise as possible. To
show that a certain claim may be true, the investigator has to show that there are some
explanations of evidence that agree with the claim. To disprove the claim, the investigator
has to show there is no explanation of evidence that agree with the
claim~\cite{printer-case}.

The authors of the FSA approach did a proof-of-concept implementation of the proposed algorithms
in CMU Common LISP~\cite{printer-case} that we target to improve the usability of by re-writing it
in a Lucid dialect,
that we call {\flucid}
(with a near-future possibility to construct a data-flow graph-based~\cite{yimin04,flucid-dfg-viz-pst2011}
IDE for the investigator to use and train novice investigators as an expert system).

In this particular work we focus on
the specification of the mentioned sample investigation case in {\flucid}
while illustrating relates fundamental concepts, operators, and application
of context-oriented case modeling and evaluation.
Common {\lisp}, unlike {\lucid}, entirely
lacks contexts build into its logic, syntax, and semantics, thereby making the
implementation of the cases more clumsy and inefficient (i.e. highly sequential).
Our system~\cite{gipsy-all-named,gipsy2005,gipsy-multi-tier-secasa09,mokhovmcthesis05,ji-yi-mcthesis-2011} (not discussed here)
offers distributed demand-driven evaluation of Lucid
programs in a more efficient way and is more general than {\lisp}'s compiler and run-time environment.

\section{Background and Related Work}

To remain stand-alone and self-sufficient in this work we recite
some material in part that we extend
from, or, deemed otherwise relevant works, such as previously
presented posters, works-in-progress, and conference papers~\cite{flucid-credibility-wips,%
flucid-blackmail-hsc09,%
self-forensics-flucid-road-vehicles,%
flucid-imf08,%
flucid-isabelle-techrep-tphols08%
,marf-into-flucid-cisse08%
} and other related cited works for the benefit of the readers.

\subsection{Intensional Logic and Programming}
\index{intensional!logic}
\index{intensional!programming}

\paragraph*{Definitions}

Intensional programming is based on intensional (or, in other words, multidimensional)
logics\index{intensional!logic}\index{logic!intensional}, which, in turn, are based on Natural Language Understanding
(aspects, such as, time, belief, situation, direction, etc.).
Intensional programming brings in {\em dimensions}\index{dimensions} and {\em context}\index{context}
to programs (e.g. space and time
in physics or chemistry). Intensional logic adds dimensions to logical
expressions; thus, a non-intensional logic\index{logic!non-intensional} can be seen as a constant or a
snapshot in all possible dimensions. {\em Intensions are dimensions} at which a
certain statement is true or false (or has some other than a Boolean value).
{\em Intensional operators}\index{intensional!operators} are operators that allow us to navigate within these
dimensions~\cite{paquetThesis}.

\paragraph*{An Example of Using Temporal Intensional Logic}
\index{logic!temporal}

Temporal intensional logic is an extension of temporal logic that allows
to specify the time in the future or in the past~\cite{paquetThesis}.

\begin{enumerate}
	\item 
$E_1$ := it is raining {\bf here} {\bf today}

Context: \{\texttt{place:}{\bf here}, \texttt{time:}{\bf today}\}
	\item 
$E_2$ := it was raining {\bf here} {\it before}({\bf today}) = {\it yesterday}
	\item 
$E_3$ := it is going to rain {\it at} (altitude {\bf here} + 500 m) {\it after}({\bf today}) = {\it tomorrow}
\end{enumerate}

The context is a collection of the dimensions, e.g. as in $E_1$'s
\api{place} and \api{time} with the corresponding tag values of {\bf here} and {\bf today}.
If we fix {\bf here} to {\bf City1} and assume it is a {\it constant}.
In the month of March, 2011, with granularity of day, for every day, we can
evaluate $E_1$ to either {\it true} or {\it false},
as shown in \xf{fig:tag-values-1d-example}.
\begin{figure}[htb!]
\hrule
\small
\centering
\begin{verbatim}

Tags days in March:   1 2 3 4 5 6 7 8 9 ...
Values (raining?):    F F T T T F F F T ...
\end{verbatim}
\normalsize
\hrule
\caption{1D Example of Tag-Value Contextual Pairs}
\label{fig:tag-values-1d-example}
\end{figure}
If one starts varying the {\bf here} dimension (which could even be broken down
to $X$, $Y$, $Z$), one gets a two-dimensional (or 4D) evaluation of $E_1$,
as shown in \xf{fig:tag-values-2d-example}.
\begin{figure}[htb!]
\hrule
\small
\centering
\begin{verbatim}

Place/Time  1 2 3 4 5 6 7 8 9 ...
City1       F F T T T F F F T ...
City2       F F F F T T T F F ...
City3       F T T T T T F F F ...
\end{verbatim}
\normalsize
\hrule
\caption{2D Example of Tag-Value Contextual Pairs}
\label{fig:tag-values-2d-example}
\end{figure}
Even with these toy examples we can immediately illustrate the hierarchical notion
of the dimensions in the context: so far the place and time we treated as
atomic values fixed at days and cities.
In some cases, we need finer
subdivisions of the context evaluation, where e.g. time can become fixed at
hour, minute, second and finer values, and so is the place broken down
into boroughs, regions, streets, etc. and finally the $X,Y,Z$ coordinates
in the Euclidean space with the values of millimeters or finer. This notion
becomes more apparent and important in {\flucid}
for evidence composition where the temporal components can be
e.g. log entries and other registered events and observations
from multiple sources.

\subsection{Lucid Overview}

{\lucid}~\cite{lucid76,lucid77,lucid85,lucid95,nonprocedural-iterative-lucid-77}
is a dataflow intensional and functional programming language. In fact, it is a
family of languages that are built upon intensional logic
(which in turn can be understood as a multidimensional generalization of temporal logic)
promoting context-aware demand-driven
parallel computation model~\cite{flucid-imf08}. A program written in some {\lucid} dialect is an expression
that may have subexpressions that need to be evaluated at certain {\em context}.
Given the set of dimensions $D=\{dim_i\}$ in which an expression varies,
and a corresponding set of indexes, or, {\it tags}, defined as placeholders
over each dimension, the context is represented as a set of $<\!\!dim_i:tag_i\!\!>$
mappings. Each variable in {\lucid}, called often a {\em stream}, is evaluated in
that defined context that may also evolve using context
operators~\cite{gipsy-simple-context-calculus-08,kaiyulucx,tongxinmcthesis08,wanphd06}.

The first generic version of {\lucid}, the General Intensional Programming Language ({\gipl})~\cite{paquetThesis},
defines two basic operators @ and \# to navigate (switch and query) in the context space $\mathcal{P}$.
The {\gipl} is the first\footnote{The second is
{\lucx}~\cite{kaiyulucx,gipsy-simple-context-calculus-08,wanphd06}, and third is TransLucid \cite{eager-translucid-secasa08}}
generic programming language of all intensional languages, defined by the means
of only two intensional operators \texttt{@} and \texttt{\#}. It has been proven that
other intensional programming languages of the Lucid family
can be translated into the {\gipl}~\cite{paquetThesis}.

\paragraph*{JLucid, Objective Lucid, and JOOIP}

{\jlucid}~\cite{mokhovmcthesis05,mokhovjlucid2005} was a first attempt on intensional arrays and ``free Java functions'' in the {\gipsy}. The approach used the {\lucid} language as the driving main computation, where Java methods were peripheral and could be invoked from the Lucid part, but not the other way around. This was the first instance of hybrid programming within the {\gipsy}. The semantics of this approach was not completely defined, plus, it was only one-sided view (Lucid-to-Java) of the problem. {\jlucid} did not support objects of any kind, but introduced the wrapper class idea for the
free Java methods.
{\olucid}~\cite{mokhovmcthesis05,mokhovolucid2005} was a logical extension of the {\jlucid} language mentioned in the previous section that inherited all of the {\jlucid}'s features and introduced Java objects to be available for use by Lucid. {\olucid} expanded the notion of the Java object (a collection of members of different types) to the array (a collection of members of the same type) and first introduced the dot-notation in the syntax and operational semantics in {\gipsy}. Like in {\jlucid}, {\olucid}'s focus was on the Lucid part being the ``main'' program and did not allow Java to call intensional functions or use intensional constructs from within a Java class. {\olucid} was the first in {\gipsy} to introduce the more complete operational semantics of the hybrid OO intensional language.
{\jooip}~\cite{gipsy-jooip-07} greatly complements {\olucid} by allowing {\java} to call the intensional language constructs closing the gap and making {\jooip} a complete hybrid OO intensional programming language within the {\gipsy} environment. {\jooip}'s semantics further refines in a greater detail the operational semantics rules of {\lucid} and {\olucid} in the attempt to make them complete.
We eliminate the OO-related aspects from this work as well as some others to conserve space
and instead focus on the context hierarchies, syntax, and semantics.

\subsection{Forensic Lucid}
\label{sect:flucid}
\label{sect:forensic-lucid}

This section summarizes
concepts and considerations in the design of the {\flucid} language,
large portions of which were studied in the earlier
work~\cite{flucid-isabelle-techrep-tphols08,flucid-imf08}.
The end goal of the language design is to
define
its constructs to concisely
express cyberforensic evidence as a context of evaluations, which can be the initial state of the case
(e.g. initial printer state when purchased from the manufacturer, see \xs{sect:printer-case-flucid}),
towards what we have actually observed (as corresponding to the final state in the Gladyshev's FSM)
(e.g. when an investigator finds the printer with two queue entries $(B_{deleted},B_{deleted})$).
One of the evaluation engines
(a topic of another work)
of the implementing
system~\cite{gipsy-all-named}
is designed to backtrace intermediate results
to provide the corresponding event reconstruction path if it exists.
The result of the expression in its basic form is either {\em true} or {\em false},
i.e. ``guilty'' or ``not guilty'' given the evidential evaluation context per explanation with the backtrace(s). There
can be multiple backtraces, that correspond to the explanation of the evidence (or lack thereof)~\cite{flucid-imf08}.

\subsubsection{Language Characteristics}
\index{{\flucid}!Properties}

We use {\flucid} to model the evidential statements and other expressions representing the evidence
and observations as context. An execution trace of a running {\flucid} program is designed to expose
the possibility of the proposed claim with the events that lead to a conclusion. {\flucid}
capitalizes its design by aggregating the features of multiple Lucid dialects
mentioned earlier
needed for these tasks along with its own extensions~\cite{flucid-imf08}.

The addition of the context calculus from {\lucx} (stands for ``Lucid enriched with context'' that
promotes contexts as first-class values) for operators on simple contexts and
context sets (\api{union}, \api{intersection}, etc.) are used to manipulate complex hierarchical context spaces
in {\flucid}. Additionally, {\flucid} inherits many of the properties of {\olucid} and
{\jooip} (Java-embedded Object-Oriented Intensional Programming language)
for the arrays and structural representation of data for modeling
the case data structures such as events, observations, and groupings
and correlation of the related data, and so on~\cite{flucid-imf08}.
Hierarchical contexts in {\flucid} also follow the example
of MARFL~\cite{marfl-context-secasa08} using a dot operator
and by overloading both @ and \# to accept different types as their
arguments.

The syntax and the operational semantics of {\flucid} were primarily maintained to be
compatible with the basic {\lucx} and {\gipl}~\cite{flucid-imf08}.
This helpful (but not absolutely necessary) when complying
with the compiler and and the runtime subsystems within the implementing system,
the General Intensional Programming System (GIPSY)~\cite{gipsy2005,gipsy-all-named}.
The translation rules or equivalent are to be provided when implementing
the language compiler within {\gipsy}, and such that the run-time environment
(General Eduction Engine, or {\gee}) can execute it with minimal changes to {\gee}'s implementation.

\subsubsection{Context of Evaluation}
\index{Context}
\index{{\flucid}!Context}

{\flucid} provides an ability to encode the ``stories'' told by
the evidence and witnesses. This constitutes the primary context
of evaluation. The ``return value'' of the evaluation is a collection
of backtraces (may be empty), which contain the ``paths of truth''. If
a given trace contains all truths values, it's an explanation
of a story. If there is no such a path, i.e. the trace, there
is no enough supporting evidence of the entire claim to be
true~\cite{flucid-imf08}.

In its simplest form, the context
can be expressed as integers or strings, to which we attribute some meaning or description.
E.g. ``1'' means printer state ``x'', etc.
The context spaces are finite and can be navigated through in all directions
of the along dimension indexes, potentially allowing negative tags
in our tag sets of dimensions.
Our contexts can also be a finite set of symbolic labels
and their values that can be internally enumerated~\cite{flucid-imf08}. The symbolic approach
is naturally more appropriate for humans and we have a machinery
to so in {\lucx}'s implementation in {\gipsy}~\cite{tongxinmcthesis08}.

We define streams of observations $os$ as our fundamental context units,
that can be a simple context or a context set. In fact, in {\flucid} we are
defining higher-level dimensions and lower-level dimensions. The highest-level
one is the {\em evidential statement} $es$, which is a finite unordered
set of observation sequences $os$.
The {\em observation sequence} $os$ is a finite {\em ordered} set of observations $o$.
The {\em observation} $o$ is an ``eyewitness'' of a particular property along
with the duration of the observation. As in the Gladyshev's
FSA~\cite{blackmail-case,printer-case} that we model after,
the basic observations are tuples of $(P,min,opt)$ in their generic form.
The observations in this form, specifically, the property $P$, can be exploded further into
{\lucx}'s context set and further into an atomic simple context~\cite{wanphd06,gipsy-simple-context-calculus-08,tongxinmcthesis08}.
(Actually $P$ can be any arbitrary expression $E$).
Context switching between different observations is done naturally with
the traditional {\lucid} @ context switching operator~\cite{flucid-imf08}.

Consider some conceptual expression of a storyboard in \xl{list:story-board-expression}
where anything in \verb+[ ... ]+ represents a story, i.e. the context of evaluation.
\texttt{foo} can be evaluated at multiple contexts (stories),
producing a collection of final results (e.g. {\em true} or {\em false}) for each story
as well as a collection of traces.

\begin{lstlisting}[
    label={list:story-board-expression},
    caption={Intensional Storyboard Expression},
    style=codeStyle
    ]
foo @
{
  [ final observed event, possible initial observed event ],
  [            ],
  [            ]
}
\end{lstlisting}

While the \verb+[...]+ notation here may be confusing with respect to the notation of \texttt{[dimension:tag]}
in traditional {\lucid} and more specifically in {\lucx}~\cite{wanphd06,tongxinmcthesis08,gipsy-simple-context-calculus-08},
it is in fact a simple syntactical extension to allow higher-level groups of contexts where this syntactical
sugar is later translated to the baseline context constructs.
The tentative notation of \verb+{[...],...,[...]}+ implies a notion similar to the notion of the ``context set''
in~\cite{wanphd06,tongxinmcthesis08} except with the syntactical sugar mentioned earlier where we allow
syntactical grouping of properties, observations, observation sequences, and evidential statements as our context sets.

The Gladyshev's concept of a generic observation sequence~\cite{printer-case}
can be expanded into the context stream using the $min$ and $opt$ values, where
they will translate into index values. Thus, $obs=(A,3,0)(B,2,0)$ expands
the property labels $A$ and $B$ into a finite stream of five indexed
elements: $AAABB$. Thus, a {\flucid} fragment in \xl{list:duplicate-context-value-tags-code} would return the third
$A$ of the $AAABB$ context stream in the observation portion of $o$. Therefore, possible evaluations to check for
the properties can be as shown in \xf{fig:duration-eval}~\cite{flucid-imf08}.

\begin{lstlisting}[
    label={list:duplicate-context-value-tags-code},
    caption={Observation Sequence With Duration},
    style=codeStyle
    ]
// Give me observed property at index 2 in the observation sequence obs
o @.obs 2
where
  // Higher-level dimension in the form of (P,min,opt)
  observation o;
  // Equivalent to writing = { A, A, A, B, B };
  // Equivalent to writing = A fby A fby A fby B fby B fby eod;
  observation sequence obs = (A,3,0)(B,2,0);
  where
    // Properties A and B are arrays of computations
    // or any Expressions
    A = [c1,c2,c3,c4];
    B = E;
    ...
  end;
end;
\end{lstlisting}

The property values of $A$ and $B$ can be anything that context calculus allows or even
more generally any arbitrary $E$ allowing to encode all kinds of case knowledge.
The \api{observation sequence} is a finite \api{ordered} context tag set~\cite{tongxinmcthesis08} that
allows an integral ``duration'' of a given tag property. This may seem like we allow duplicate tag values that are
unsound in the classical {\lucid} semantics; however, we find our way around it
with the implicit tag index. The semantics of the arrays of computations is not a part of either
{\gipl} or {\lucx}; however, the arrays are provided by
{\olucid}. We use the notion
of the arrays to evaluate multiple computations at the same context. Having an array of computations
is conceptually equivalent of running an a {\lucid} program under the same context for each array element
in a separate instance of the evaluation engine and then the results of those expressions are gathered
in one ordered storage within the originating program. Arrays in {\flucid} are needed to represent
a set of results, or {\em explanations} of evidential statements, as well as denote some properties
of observations.
(We
explore the notion of arrays in {\flucid} much greater detail in
another work)~\cite{flucid-imf08}.

To make equivalence relation with the formal Gladyshev's FSA approach,
computations $c_i$ correspond to the states $q$ and event $i$ that
enable the transition. For {\flucid}, we define
$c_i$ as theoretically any Lucid expression $o = E$~\cite{flucid-imf08}.

\begin{figure}[htb!]
\hrule
\small
\scriptsize
\begin{verbatim}

Observed property (context): A A A B B
        Sub-dimension index: 0 1 2 3 4

o @.obs 0 = A
o @.obs 1 = A
o @.obs 2 = A
o @.obs 3 = B
o @.obs 4 = B

To get the duration/index position:

o @.obs A = 0 1 2
o @.obs B = 3 4
\end{verbatim}
\normalsize
\hrule
\vspace{5pt}
\caption{Handling Duration of an Observed Property in the Context}
\label{fig:duration-eval}
\end{figure}

In \xf{fig:duration-eval} a possibility is illustrated to query for the sub-dimension indices by raw property
where it is present. This produces a finite stream of valid indices that can be used in subsequent expressions, or,
alternatively by supplying the index we can get the corresponding raw property at that index. The latter
feature is still under investigation of whether it is safe to expose it to {\flucid} programmers or make
it implicit at all times at the implementation level.
This method of indexing was needed to remedy the ``problem'' of ``duplicate tags'':
as previously mentioned, observations form the context and allow durations. This means multiple
duplicate dimension tags with implied subdimension indexes should be allowed as
the semantics of traditional Lucid approaches do not allow duplicate
dimension tags. It should be noted however, that the combination of
the tag and its index in the stream is still unique and is nicely folded
into the traditional Lucid semantics~\cite{flucid-imf08}.

\subsubsection{Transition Function}
\label{sect:tans-func}
\index{Transition Function}
\index{{\flucid}!Transition Function}

A transition function (derived from the same notion from the works of Gladyshev et al.~\cite{printer-case,blackmail-case})
determines how the context of evaluation changes
during computation. It represents in part the case's crime scene modeling.
A general issue exists that we have to address
is that the transition function {\trans} is usually problem-specific.
In the FSA approach, the transition function is the labeled graph itself \cite{printer-case}.
We follow the graph of the case to model our {\flucid} equivalent~\cite{flucid-imf08}.

In general,
{\lucid} has already basic operators to navigate and switch from one context to another,
that can be said equivalent to state switching.
These operators represent the basic ``built-in'' transition functions in themselves (the intensional
operators such as @, \#, \api{iseod}, \api{first}, \api{next}, \api{fby}, \api{wvr}, \api{upon},
and \api{asa} as well as their inverse operators~\cite{flucid-imf08}.
However, a specific problem being modeled requires more specific transition function
than just plain intensional operators. In this case the transition function is
a {\flucid} function where the matching state transition modeled through
a sequence of intensional operators~\cite{flucid-imf08}.
In fact, the forensic operators are just pre-defined functions that rely on the traditional
and inverse Lucid operators as well as context switching operators
that achieve something similar to the transitions~\cite{flucid-imf08}.
At the implementation level, it is the {\gee} that actually does
the execution of {\trans} within {\gipsy}.
In fact, the intensional operators
of {\lucid} represent the basic building blocks for {\trans} and {\invtrans}.

\subsubsection{Generic Observation Sequences}
\index{Generic Observation Sequences}
\index{{\flucid}!Generic Observation Sequences}

We adopt a way of modeling generic observation sequences as an equivalent to the \api{box}
operator from the {\lucx}'s context calculus~\cite{wanphd06,tongxinmcthesis08}
in the dimensional context that defines the space of all possible evaluations.
The generic observation sequence context contains observations whose properties' duration
is not fixed to the $min$ value as in $(P,min,0)$ as we studied so far. The third position
in the observation tuple, $opt$ is not $0$ in the generic observation and as a result in
the containing observation sequence, e.g. $os=(P_1,1,2)(P_2,1,1)$.
Please refer to~\cite{printer-case,blackmail-case,flucid-imf08,flucid-isabelle-techrep-tphols08}
for more detailed
examples of a generic observation sequence~\cite{flucid-imf08}.

\subsubsection{Primitive Operators}
\index{Forensic Lucid!New Operators}
\index{Forensic Lucid!Operators}
\index{Operators!Forensic Lucid}

The basic set of the classic intensional operators is extended with the similar
operators, but inverted in one of their aspects: either negation of trueness or reverse of direction of navigation.
Here we provide
a definition of these operators alongside with the classical ones (to
remind the reader what they do and enlighten the unaware reader). The reverse operators have a restriction that they must work on
the bounded streams at the positive infinity. This is not a stringent limitation as the
our contexts of observations and evidence in this work are always finite, so they all have
the beginning and the end. What we need is an ability to go back in the stream and, perhaps,
negate in it with classical-like operators, but reversed~\cite{flucid-imf08}.
\index{Forensic Lucid!Operators @ and \#}

Following the steps in~\cite{paquetThesis}, we further represent the definition of the
operators via @ and \#.
Again, there is a mix of classical operators that were previously
defined in \cite{paquetThesis}, such as \lucidop{first}, \lucidop{next},
\lucidop{fby}, \lucidop{wvr}, \lucidop{upon}, and \lucidop{asa} as well
as the new operators from this work~\cite{flucid-imf08}.

\subsubsection{Forensic Operators}

The operators presented here are based on the discussion of the
combination \cite{printer-case} function and others that form more-than-primitive
operations to support the required implementation.
The \api{comb()} operator is
realized in the general manner in {\flucid}
for combining analogies of multiple partitioned runs (MPRs) \cite{printer-case},
which in our case are higher-level contexts, in the new language's dimension types~\cite{flucid-imf08}.

\begin{itemize}

\item
\lucidop{combine} corresponds to the $comb$ function described earlier.
It is defined in \xl{list:combine}.

\begin{lstlisting}[
    label={list:combine},
    caption={The \lucidop{combine} Operator},
    style=codeStyle
    ]
/**
 * Append given e to each element of a given
 * stream e under the context of d.
 * @return the resulting combined stream
 */
combine(s, e, d) =
  if iseod s then eod;
  else (first s fby.d e) fby.d combine(next s, e, d);
\end{lstlisting}

\item
\lucidop{product}
corresponds to the cross-product of contexts, translated from that
of the {\lisp} example and added with context. It is defined in \xl{list:product}.

\begin{lstlisting}[
    label={list:product},
    caption={The \lucidop{product} Operator},
    style=codeStyle
    ]
/**
 * Append elements of s2 to element of s1
 * in all possible combinations.
 */
product(s1, s2, d) =
  if iseod s2 then eod;
  else combine(s1, first s2) fby.d product(s1, next s2)
\end{lstlisting}

\end{itemize}

\section{Modeling Printer Case in Forensic Lucid}
\label{sect:printer-case-flucid}

\subsection{ACME Manufacturing Printing Case}
\index{Printer Case}
\index{Case!Printer}

\noindent
This is one of the cases we re-examine from the Gladyshev's FSA approach~\cite{printer-case}.

\begin{quote}{\it
The local area network at some company called ACME Manufacturing consists of two
personal computers and a networked printer. The cost of running
the network is shared by its two users Alice (A) and Bob (B). Alice,
however, claims that she never uses the printer and should not be
paying for the printer consumables. Bob disagrees, he says that he
saw Alice collecting printouts. According to the manufacturer, the
printer works as follows:

\begin{enumerate}
\item
When a print job is received from the user, it is stored in the first
unallocated directory entry of the print job directory.
\item
The printing mechanism scans the print job directory from the beginning
and picks the first active job.
\item
After the job is printed, the corresponding directory entry is marked as
``deleted'', but the name of the job owner is preserved.
\item
The printer can accept only one print job from each user at a time.
\item
Initially, all directory entries are empty.
\end{enumerate}

\noindent
The investigator finds the current state of the printer's buffer as:

\begin{enumerate}
\item Job From B Deleted
\item Job From B Deleted
\item Empty
\item Empty
\item ...
\end{enumerate}
}
\end{quote}

\paragraph{Investigative Analysis}

If Alice never printed anything, only one directory
entry must have been used, because the printer accepts
only one print job from each user at a time~\cite{printer-case}. However, two
directory entries have been used and there are no
other users except Alice and Bob. Therefore, it must
be the case that both Alice and Bob submitted their
print jobs in the same time frame. The trace of Alice's
print job was overwritten by Bob's subsequent print
jobs. As a result, a finite state machine is constructed
to model the situations as in the FSA~\cite{printer-case}
to indicate the initial state and other possible states
and how to arrive to them when Alice or Bob would have
submitted a job and a job would be deleted~\cite{printer-case}.
The FSM presented in \cite{printer-case} covers the entire case with all possible events and transitions
resulted due to those events. It is modeled based on the properties of
the investigation, in this case the printer queue's state according to the
manufacturer specifications and the two potential users. The modeling is assumed
to be done by the investigator in the case in order to perform a thorough analysis.
It also doesn't really matter how actually it so happened that the Alice's print job
was overwritten by Bob's subsequent jobs as is not a concern for this case any further.
Assume, this behavior is derived from the manufacturer's specification and the evidence found.
The investigator will have to make similar assumptions in the real case~\cite{printer-case}.

The authors of~\cite{printer-case} provided a proof-of-concept implementation of
this case in Common LISP (not recited in here)
which takes about 6-12 pages of printout depending on the printing options set and column format.
Using our proposed solution, we rewrite the example in {\flucid} and show the advantages of a much finer
conciseness and added benefit of the implicit context-driven expression and evaluation,
and parallel evaluation that the LISP
implementation lacks entirely.

\subsection{Sample Forensic Lucid Specification}

The simulated printer case is specified in {\flucid} as follows.
{\trans} is implemented in \xl{list:acme-flucid-transition-function}.
We then provide the
implementation of {\invtrans} in~\cite{flucid-imf08}
in \xl{list:acme-flucid-inv-transition-function}.
Finally, the ``main program'' is modeled in \xl{list:flucid-claims-macros}
that sets up the context hierarchy and the invokes {\invtrans}.
This specification is the translation of the LISP implementation
by Gladyshev described earlier \cite{printer-case}
and described in this section in semi-structured English.

\subsubsection*{The ``Main Program''}

In \xl{list:flucid-claims-macros} where the computation begins
in our {\flucid} example.
This is an equivalent of \api{main()} or program entry point
in other mainstream languages. The goal of this fragment
is to setup the context of evaluation which is core to the
case -- the evidential statement \api{es}. This is
the highest level dimension in Lucid terms, and it is
hierarchical. This is an unordered list (set) of stories
and witness accounts of the incident (themselves known
as observation sequences); ordering in the
program of them is arbitrary and has an array-like
structure. The relevant stories to the incident are that
of Alice, the evidence of the printer's final state
as found by the investigator, and the ``expert testimony''
by the manufacturer of how the printer works. These observation
sequences are in turn defined as ordered collections of
observations nesting one lever deeper into the context.
The printer's final state dimension \api{F} is the only
observation for the printer found by the investigator,
which is an observation of the property of the
printer's queue ``Bob's job deleted last'' syntactically
written as ``B\_deleted'' as inherited from Gladyshev's
notation. Its duration is nothing special, that it was
simply present. The \api{manuf} observation sequence
dictated by the manufacturer's specification that the
printer's queue state was empty initially for an undetermined
period \api{\$} of time when the printer was delivered.
These are two observations, followed in time/
Alice's line (also tow observations) is that from the beginning Alice did not
not any actions signified by the properties $P$ such as 
``add\_B'' or ``take'' (implying the computation ``add\_A''
has never happened (0 duration for the ``infinity'' i.e. till
the investigator examined the printer); which is Alice's claim.
\api{alice\_claim} is a collection of Boolean results for possible explanations
or lack thereof for Alice's claim in this case at the
context of all this evidence and as evaluated by \api{invpsiacme} {\invtrans}.
If Alice's claim were to check out; the results would be ``true'';
``false'' otherwise.

\begin{lstlisting}[
    label={list:flucid-claims-macros},
    caption={Developing the Pinter Case: ``main''},
    style=codeStyle
    ]
alice_claim @ es
where
  evidential statement es = [ printer, manuf, alice ];

  observation sequence printer = F;
  observation sequence manuf = [Oempty, $];
  observation sequence alice = [Oalice, F];

  observation F = (``B_deleted'', 1, 0);
  observation Oalice = (P_alice, 0, +inf);
  observation Oempty = (``empty'', 1, 0);

  // No ``add_A''
  P_alice = unordered {``add_B'', ``take''};

  alice_claim = invpsiacme(F, es);
end;
\end{lstlisting}

\subsubsection*{Modeling Forward Transition Function {\trans}}

In \xl{list:acme-flucid-transition-function} {\trans} illustrating
the normal flow of operations to model the scene. Which is also
a translation from LISP from Gladyshev \cite{printer-case} using
{\flucid} syntax and operators described in \cite{flucid-imf08}.
The function is modeled per manufacturer specification and focuses
on the queue of the printer. ``A'' corresponds to ``Alice'' and ``B''
to ``Bob'' along with their prompted queue actions to add deleted
print jobs. The code is a rather straightforward translation of the
FSM/LISP code in \cite{printer-case}. \api{S} is a collection of
state properties observed. \api{c} is a ``computation'' action
to add or take print jobs by the printer's spooler. \api{d} is
a classical Lucid dimension type along which the computation is happening
(there can be multiple dimensions and evaluations going on).

\begin{lstlisting}[
    label={list:acme-flucid-transition-function},
    caption={``Transition Function'' {\trans} in {\flucid} for the ACME Printing Case},
    style=codeStyle
    ]
acmepsi(c, s, d) =
  // Add a print job from Alice
  if c == ``add_A'' then
    if d1 == ``A'' || d2 == ``A'' then s;
    else
      if d1 in S then ``A'' fby.d d2;
      else
        if d2 in S then d1 fby.d ``A'';
        else s;
  // Add a print job from Bob
  else if c == ``add_B'' then
    if d1 == ``B'' || d2 == ``B'' then s;
    else
      if d1 in S then ``B'' fby.d d2;
      else
        if d2 in S then d1 fby.d ``B'';
        else s;
  // Printer takes the job per manufacturer specification
  else if c == ``take''
    if d1 == ``A'' then ``A_deleted'' fby.d d2;
    else
      if d1 == ``B'' then ``B'' fby.d d2;
      else
        if d2 == ``A'' then d1 fby.d ``A_deleted'';
        else
          if d2 == ``B'' then d1 fby.d ``B_deleted'';
          else s;
  // Done
  else s fby.d eod;

  where
    dimension d;
    S = [``empty'', ``A_deleted'', ``B_deleted''];
    d1 = first.d s;
    d2 = next.d d1;
  end;
\end{lstlisting}

\subsubsection*{Modeling Inverse Transition Function {\invtrans}}

In \xl{list:acme-flucid-inv-transition-function} is the inverse {\invtrans}
backtracking implementation with the purpose of event reconstruction,
also translated from {\lisp} to {\flucid} like the preceding fragments
using the {\flucid} operators. It is naturally more complex than
{\trans} due to a possibility of choices (non-determinism) when going back 
in time so all of them have to be explored. This backtracking, if successful,
for any claim, would provide the Gladyshev's ``explanation'' of that claim,
i.e. the claim attains its meaning and is validated within the provided
evidential statement. {\invtrans} is based on the traversal from \api{F}
to the initial observation of the printer's queue as defined in ``main''.
If such path were to exist, then Alice's claim would have had an explanation.
\api{pby} ({\em preceeded by}) is the {\flucid} inverse operator of classical
Lucid's \api{fby} ({\em followed by}). \api{backtraces} is an array of event
backtracing computations identified with variables; their number and definitions
depend on the crime scene and are derived from the state machine of Gladyshev.

\begin{lstlisting}[
    label={list:acme-flucid-inv-transition-function},
    caption={``Inverse Transition Function'' {\invtrans} in {\flucid} for the ACME Printing Case},
    style=codeStyle
    ]
invpsiacme(s, d) = backtraces
where
  backtraces = [A, B, C, D, E, F, G, H, I, J, K, L, M ];
  where
    A = if d1 == ``A_deleted''
        then d2 pby.d ``A'' pby.d ``take'' else eod;

    B = if d1 == ``B_deleted''
        then d2 pby.d ``B'' pby.d ``take'' else eod;

    C = if d2 == ``A_deleted'' && d1 != ``A'' && d2 != ``B''
        then d1 pby.d ``A'' pby.d ``take'' else eod;

    D = if d2 == ``B_deleted'' && d1 != ``A'' && d2 != ``B''
        then d1 pby.d ``B'' pby.d ``take'' else eod;

    E = if d1 in S && d2 in S
        then s pby.d ``take'' else eod;

    F = if d1 == ``A'' && d2 != ``A''
        then
          [ d2 pby.d ``empty'' pby.d ``add_A'',
            d2 pby.d ``A_deleted'' pby.d ``add_A'',
            d2 pby.d ``B_deleted'' pby.d ``add_A'' ]
        else eod;

    G = if d1 == ``B'' && d2 != ``B''
        then
          [ d2 pby.d ``empty'' pby.d ``add_B'',
            d2 pby.d ``A_deleted'' pby.d ``add_B'',
            d2 pby.d ``B_deleted'' pby.d ``add_B'' ]
        else eod;

    H = if d1 == ``B'' && d2 == ``A''
        then
          [ d1 pby.d ``empty'' pby.d ``add_A'',
            d1 pby.d ``A_deleted'' pby.d ``add_A'',
            d1 pby.d ``B_deleted'' pby.d ``add_A'' ]
        else eod;

    I = if d1 == ``A'' && d2 == ``B''
        then
          [ d1 pby.d ``empty'' pby.d ``add_B'',
            d1 pby.d ``A_deleted'' pby.d ``add_B'',
            d1 pby.d ``B_deleted'' pby.d ``add_B'' ]
        else eod;

    J = if d1 == ``A'' || d2 == ``A''
        then s pby.d ``add_A'' else eod;

    K = if d1 == ``A'' && d2 == ``A''
        then s pby.d ``add_B'' else eod;

    L = if d1 == ``B'' && d2 == ``A''
        then s pby.d ``add_A'' else eod;

    M = if d1 == ``B'' || d2 == ``B''
        then s pby.d ``add_B'' else eod;

    where
      dimension d;
      S = [``empty'', ``A_deleted'', ``B_deleted''];
      d1 = first.d s;
      d2 = next.d d1;
    end;
\end{lstlisting}

\section{Conclusion}

We presented the basic overview of {\flucid}, its concepts, ideas,
and dedicated purpose -- to model, specify, and evaluation digital
forensics cases. The process of doing so is significantly simpler
and more manageable than the previously proposed FSM model and its
common LISP realization. At the same time, the language is founded
in more than 30 years research on correctness and soundness of 
programs and the corresponding mathematical foundations of the
{\lucid} language, which is a significant factor should a {\flucid}-based
analysis be presented in court. We re-wrote in {\flucid} one of the sample cases
initial modeled by Gladyshev in the FSM and Common LISP to show the
specification is indeed more manageable and comprehensible than
the original and fits in two pages in this paper.

We also still realize by looking at the examples the usability aspect is
still desired to be improved further for the investigators, especially
when modeling {\trans} and {\invtrans}, as a potential limitation, prompting
one of the future work items to address it further.

In general, the proposed practical approach in the cyberforensics field
can also be used to model and evaluate normal investigation process involving
crimes not necessarily associated with information technology.
Combined with an expert system (e.g. implemented in CLIPS~\cite{clips}),
it can also be used in training new staff in investigation techniques.
The
notion of
hierarchical contexts as first-class values brings more understanding
of the process to the investigators in cybercrime case management tools.

\section{Future Work}

\begin{itemize}
\item
	Formally prove equivalence to the FSA approach.
\item
	Adapt/re-implement a graphical UI based on the data-flow graph tool~\cite{yimin04} to
	simplify {\flucid} programming further for not very tech-savvy
	investigators by making it visual. The listings provided are not
	very difficult to read and quite manageable to comprehend, but any
	visual aid is always an improvement.
\item
	Refine the semantics of {\lucx}'s context sets and their operators to be
	more sound, including Box and Range.
\item
	Explore and exploit the notion of credibility factors of the
	evidence and witnesses fully.
\item
	Release a full standard {\flucid} specification.
\end{itemize}

\section*{Acknowledgments}

This research work was funded by NSERC and the Faculty
of Engineering and Computer Science of Concordia University,
Montreal, Canada.
We would also like to acknowledge the reviewers who took time to
do a constructive quality review of this work.
Thanks to Andrei Soeanu for presenting this work at ICDF2C 2011
on behalf of the authors.

\label{sect:bib}
\bibliographystyle{abbrv}
\bibliography{flucid-printer-case-arXiv}

\printindex

\end{document}